\documentclass[letterpaper, 10 pt, conference]{ieeeconf} \IEEEoverridecommandlockouts       
\usepackage{cite}
\usepackage{amsmath,amssymb,amsfonts}
\usepackage{algorithmic}
\usepackage{graphicx}
\usepackage{textcomp}
\usepackage{xcolor}
\usepackage{subfigure}
\def\BibTeX{{\rm B\kern-.05em{\sc i\kern-.025em b}\kern-.08em     T\kern-.1667em\lower.7ex\hbox{E}\kern-.125emX}}
\usepackage{balance}

\makeatletter
\let\NAT@parse\undefined
\makeatother
\usepackage{hyperref}  %hyperref still needs to be put at the end!

\hypersetup{
colorlinks=true,
linkcolor=blue,
urlcolor=blue,
citecolor=blue,
}
    
\begin{document}

\title{\LARGE \bf
LimSim: A Long-term Interactive Multi-scenario Traffic Simulator
}
\author{Licheng Wen$^{1, \dagger}$, Daocheng Fu$^{1, \dagger}$, Song Mao$^{1}$, Pinlong Cai$^{1, \ast}$, Min Dou$^{1}$, Yikang Li$^{1}$ and Yu Qiao$^{1}$% <-this % stops a space 
\thanks{$^{1}$ All authors are with the Shanghai Artificial Intelligence Laboratory, Shanghai, China, 200032, Email: 
        {\tt\small \{wenlicheng, fudaocheng, maosong, caipinlong, doumin, liyikang, qiaoyu\}@pjlab.org.cn}.}%
\thanks{$^{\dagger}$ Equal contribution; Order determined by a coin toss.}%
\thanks{$^{\ast}$ Corresponding author.}%
}

% \ast \dagger

\maketitle

\begin{abstract}

With the growing popularity of digital twin and autonomous driving in transportation, the demand for simulation systems capable of generating high-fidelity and reliable scenarios is increasing. Existing simulation systems suffer from a lack of support for different types of scenarios, and the vehicle models used in these systems are too simplistic. Thus, such systems fail to represent driving styles and multi-vehicle interactions, and struggle to handle corner cases in the dataset. In this paper, we propose LimSim, the Long-term Interactive Multi-scenario traffic Simulator, which aims to provide a long-term continuous simulation capability under the urban road network. LimSim can simulate fine-grained dynamic scenarios and focus on the diverse interactions between multiple vehicles in the traffic flow. This paper provides a detailed introduction to the framework and features of the LimSim, and demonstrates its performance through case studies and experiments. The source code of LimSim is available on GitHub: \href{https://www.github.com/PJLab-ADG/LimSim}{github.com/PJLab-ADG/LimSim}.

\end{abstract}

% \begin{IEEEkeywords}
% Traffic flow simulation, Multi-vehicle interaction, Scenario evaluation, Simulation system.

% \end{IEEEkeywords}

\section{Introduction}

The urban traffic system is complex and diverse, encompassing numerous types of roads, such as intersections, ramps, and roundabouts, as well as a wide variety of traffic participants. Vehicles are the principal component of road scenarios, and drivers with different driving styles exhibit heterogeneous behavior while driving \cite{sagberg2015review,zhang2021demystifying}. 
By providing a comprehensive view of how urban transport systems operate, traffic simulators offer policymakers valuable insights into the potential real-world impacts of their decisions \cite{t1997preparing, teo2012evaluating, daniel2021city}. 
Simulators also allow potential problems to be identified and addressed before implementation in the physical environment, resulting in improved transportation while saving time and resources.
With the wide use of advanced technologies like virtual reality and data mining, simulation  focusing on driving scenarios has gained significant attention in recent years and is becoming a crucial element in the construction of the digital twins in cities  \cite{wang2022automatic,argota2022getting}. 
Scenario-oriented simulation also plays a vital role in the development and testing of autonomous driving. It enables testing and validation of multiple algorithms and systems by conducting simulations in different scenarios, environments, system configurations, and driving characteristics \cite{schoner2018Simulation}. 

Generally, simulators can be classified into three kinds, including flow-based simulators (e.g., PARAMICS \cite{cameron1996PARAMICS}, Vissim \cite{fellendorf2010Microscopic}, Aimsun \cite{casas2010Traffic}, SUMO \cite{lopez2018Microscopic}), vehicle-based simulators (e.g., AirSim \cite{shah2017AirSim}, LGSVL \cite{rong2020lgsvl}, CARLA \cite{dosovitskiyCARLA}, MetaDrive \cite{li2022MetaDrive}), and data-based simulators (e.g., SimNet \cite{bergamini2021SimNet}, InterSim \cite{sun2022InterSim}, TrafficGen \cite{feng2022trafficgen}). 
% that depend on data collected by a fleet of vehicles to reconstruct traffic flows. 
While flow-based simulators can reflect the characteristics of the real-world transportation system and analyze various traffic flow conditions, they tend to simplify vehicle motion behaviors and fail to consider multi-vehicle interactions and kinematics constraints. 
The vehicle-based simulators aim to construct virtual road scenarios to verify the performances of developed algorithms. However, the generation of background traffic flow relies on manual editing scenarios or using collected road data, which fails to accurately reproduce the characteristics of actual scenarios.
In addition, the data-based simulators rely on real-world traffic data. By extracting vehicle motion features from the datasets, these simulators allow ego vehicle to interact with the background traffic flow. However, the datasets typically provide fragmented, small-scale scenarios, making these simulators unable to conduct long-term continuous simulations.

\begin{figure*}
    \centering
     \includegraphics[width=0.98\linewidth,trim={5cm 2cm 5cm 4.5cm},clip]{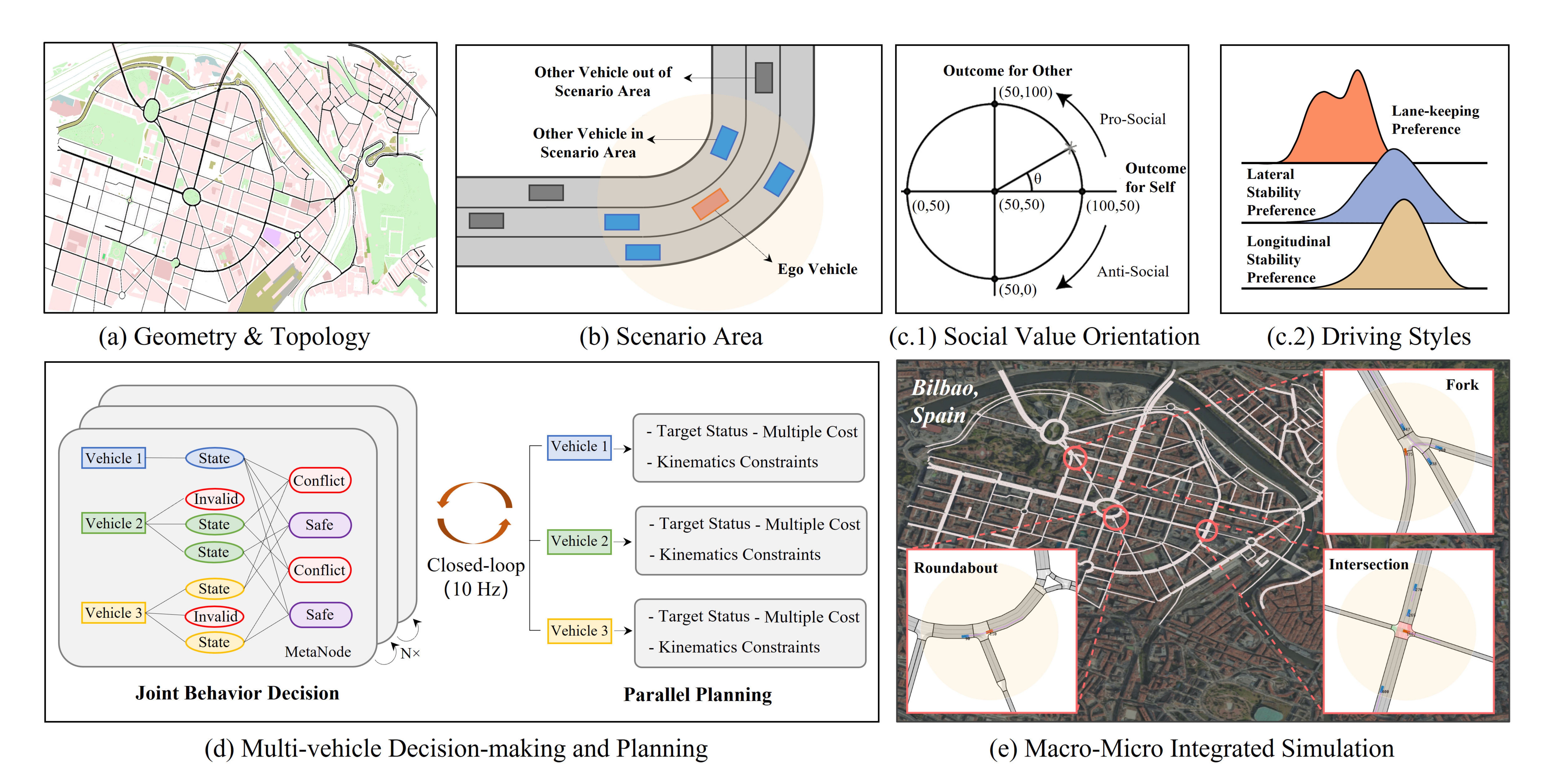}
      \caption{(a) LimSim supports the representation of both geometric information and topological structure of road networks derived from vector maps. (b) The range of microscopic simulation is defined by the scenario area. Vehicles outside this area are controlled by classic traffic flow models, while those within the range undergo precise decision-making and planning processes. (c.1) The social value orientations quantify the balance of interests in the process of multi-vehicle interaction. (c.2) Heterogeneous driving behaviors are taken into account during trajectory planning. (d) Multi-vehicle joint behavior decision-making allows all vehicles with potential interactions to display mature driving maneuvers. High-frequency parallel trajectory planning considers vehicle's kinematic constraints and ensures runtime efficiency. (e) LimSim combines macroscopic traffic flow simulations for the entire road network with microscopic vehicle motion simulations in specific scenarios, providing a comprehensive representation of traffic behaviour. }
      \label{fig:introduction}
      \vspace{-10pt}
\end{figure*}

To overcome these limitations, we propose LimSim, the Long-term Interactive Multi-scenario Traffic Simulator. 
LimSim simulates dynamic traffic scenarios in urban road networks at a fine-grained level.
It includes a multi-scenario road network construction module, a multi-source traffic flow generation module, a multi-vehicle joint decision planning module, and a multi-dimensional scenario analysis module. The core features of LimSim are listed below:

\begin{itemize}
    \item \textbf{Long-term}: Traffic flow can be generated over long periods under the guidance of demand construction and route planning. 
    Both geometric information and topological structure are essential inputs to represent a city-level road network, as shown in Fig. \ref{fig:introduction}(a).
    \item \textbf{Diversity}: The built-in behavioral models take heterogeneous social value orientations and driving styles of vehicles into account, supporting various driving behaviors like car-following, lane-changing, and merging, as shown in Fig. \ref{fig:introduction}(c.1) and \ref{fig:introduction}(c.2).
    \item \textbf{Interactivity}: Vehicles in the scenario area are controlled by a joint decision-making and planning framework, offering sophisticated interactions among vehicles, as shown in Fig. \ref{fig:introduction}(b) and \ref{fig:introduction}(d).
    \item \textbf{Multi-scenario}: The universal road components are proposed to support a variety of road structures in the real world. These components integrate macroscopic and microscopic scenarios and thus ensure a consistent simulation system, as shown in Fig. \ref{fig:introduction}(e). 
\end{itemize}

\section{Related work} 

\subsection{Flow-based simulators}
The development of the flow-based simulating systems has been going on for decades, intending to provide technical support for urban planners and managers. PARAMICS (Parallel Microscopic Simulator) was released as commercial software by Quadstone in 1998, which integrates simulation, visualization, interactive road network drawing, adaptive signal control, online simulation data statistical analysis, and traffic control strategy evaluation, among other features \cite{cameron1996PARAMICS}. 
Vissim is a commercial microscopic traffic simulation software developed by PTV GmbH, providing high-level visualization of complex traffic situations with realistic traffic models \cite{fellendorf2010Microscopic}. 
CORSIM (CORridor SIMulation)  is a traffic engineering analysis and modeling software developed with support from the Federal Highway Administration (FHWA), which can perform complex road geometry modeling, large-scale traffic condition simulation, traffic control, and management simulation, road network topology and vehicle interaction \cite{owen2000Traffic}. 
Aimsun is widely used by traffic engineers for traffic planning, microscopic traffic simulation, traffic demand, and related data analysis \cite{casas2010Traffic}. 
SUMO (Simulation of Urban MObility)  is a free and open-source traffic flow simulation software developed by the German Aerospace Center (DLR) for modeling intermodal traffic systems, including various modes of transportation, and can perform traffic network construction, route planning, emission calculations, and other functions, with multiple APIs to control the simulation remotely \cite{lopez2018Microscopic}.

While flow-based simulators are powerful in facilitating large-scale traffic simulations to demonstrate the overall traffic conditions of the road network, they generally rely on simplistic car-following and lane-changing models to control vehicles, resulting in limited vehicle behavior simulations. 
Moreover, due to the lack of strict vehicle kinematic constraints, flow-based simulators cannot accurately capture the microscopic movements of vehicles, thus reducing their simulation fidelity.

\begin{table*}[htbp]
\caption{Comparisons of existing simulators}
\begin{center}
\begin{tabular}{ccccccc}
\hline
Simulator & Long Term & Vehicle Dynamic &  Diverse Interaction  & Custom  Trajectory & Cost Efficiency & Transferability \\
\hline
Flow-based        &  $+$        &  $-$        &  $\circ$  &  $+$        &  $\circ$    &  $+$  \\
Vehicle-based     &  $\circ$    &  $+$        &  $-$      &  $\circ$    &  $-$        &  $+$  \\
Data-based    &  $\circ$    &  $\circ$    &  $\circ$  &  $-$        &  $\circ$    &  $-$  \\
\textbf{LimSim}       &  $+$        &  $+$        &  $+$      &  $+$        &  $+$        &  $+$  \\
\hline
\multicolumn{4}{l}{ $-$ Not Available \hspace{1cm}  $\circ$ Basic Performance \hspace{1cm}  $+$ Good Performance}
\end{tabular}
\label{tab1}
\end{center}
\vspace{-15pt}
\end{table*}

\subsection{Vehicle-based simulators}

Vehicle-based simulation can also provide realistic dynamic simulations \cite{xu2019Automated}. Early autonomous driving  software used modified commercial game engines for physical simulation and visualization, providing fundamental physical assets and interaction between objects \cite{craighead2007Survey}, including USARSim \cite{carpin2007usarsim}, Webots \cite{michel2004cyberbotics}, SimRobot \cite{laue2006simrobot}, Unity \cite{juliani2018unity}, etc. These autonomous driving simulation engine requires compatible open interfaces, obtainable object, and sensor parameters, realistic graphics rendering and physical engine, and deployment of cost efficiency \cite{pilz2019Development}. The following simulators are very popular nowadays.

Gazebo is an open-source, scalable, and flexible multi-robot 3D simulator, which is usually used along with ROS to simulate the physical characteristics of mock objects and realize dynamic 3D rendering and interactive communication \cite{koenig2004design}. 
AirSim is an Unreal Engine-based simulator that provides physically and visually realistic simulations for a variety of purposes, which includes vehicle simulations, city scenarios, 
APIs to simplify programming, and plug-and-play code to quickly create rich scenarios \cite{shah2017AirSim}. 
LGSVL is an open-source autonomous driving simulator developed by LG's Silicon Valley laboratory based on the Unity engine, which permits users to label and export for a high-precision map format and also provides support for multi-type sensor simulation including LiDAR, millimeter wave radar, GPS, IMU, and camera \cite{rong2020lgsvl}. 
CARLA is an open-source simulator for autonomous driving research, which provides open digital assets (including urban layouts, buildings, vehicles, etc.) and supports flexible specifications of sensor suites and environmental conditions \cite{dosovitskiyCARLA}. 
MetaDrive is an easy-to-install and powerful driving simulation platform that is used to explore universal reinforcement learning algorithms to support machine autonomy. It generates an infinite number of driving scenarios from generating programs and importing real data \cite{li2022MetaDrive}. 

The vehicle-based simulators have more precise kinematic constraints for trajectory planning, providing a more realistic driving environment for testing autonomous driving decision and planning algorithms. However, this type of simulations lacks interactive modeling of vehicles, and cannot provide background traffic flow with real scenario characteristics.

\begin{figure*}[tp]
    \centering
     \includegraphics[width=0.80\linewidth]{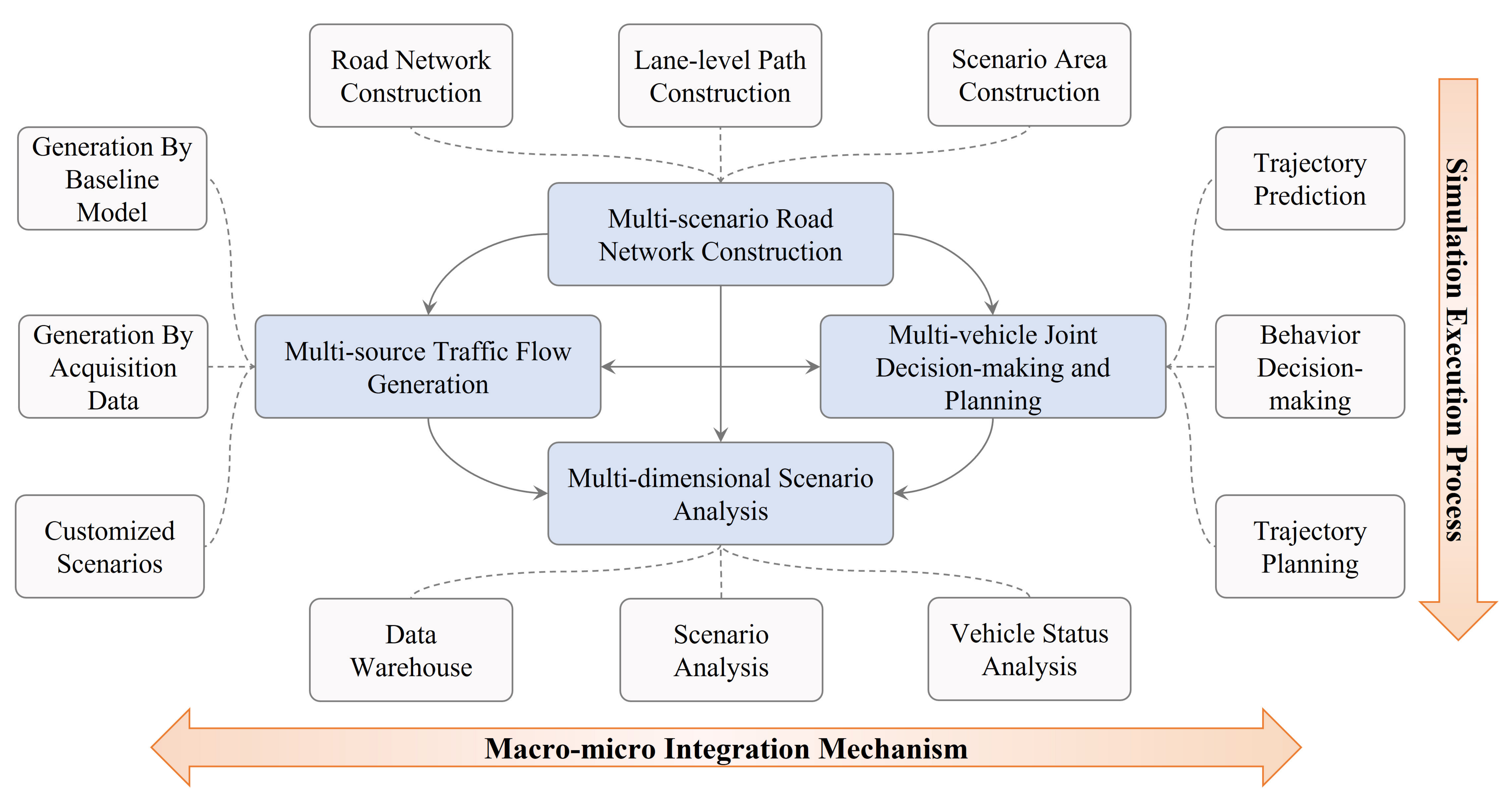}
      \caption{Framework of LimSim. LimSim comprises a multi-scenario road network construction module, a multi-source traffic flow generation module, a multi-vehicle joint decision planning module, and a multi-dimensional scenario analysis module.}
      \label{fig:framework}
\vspace{-15pt}
\end{figure*}

\subsection{Data-based simulators}
Simulators that use traffic flow data appeals to many researchers. 
During the simulation process, the ego vehicles will take different behaviors from the dataset, making it difficult to consider the dynamic interaction between ego vehicles and other vehicles in the road environments, since other vehicles need to respond the behavior of ego vehicle.
This type of work is called data-based simulators, and some of the latest work are summarized as follows.

SimNet is the first to present a machine learning-based simulation system for ego vehicles that generates realistic driving episodes based on historical driving data \cite{bergamini2021SimNet}. The system's performance improves with the amount of data used for training and can help identify issues in planning systems. The approach is more scalable than hand-coding actors and can achieve high realism by directly learning behavior from observed data. 
InterSim is an interactive traffic simulator used to test autonomous driving planners, which infers the interaction relationships between vehicles in the scenarios and generates realistic trajectories consistent with these relationships for environmental vehicles \cite{sun2022InterSim}. 
TrafficGen is also a data-driven method for generating realistic traffic scenarios. It uses an autoregressive generative model with an encoder-decoder architecture to synthesize complete traffic scenarios.
Specifically, TrafficGen places vehicles on a given map and uses a motion forecasting model to generate multi-mode diverse samples for vehicles' trajectories \cite{feng2022trafficgen}. 

Data-based simulators can implicitly learn multi-vehicle interactions from natural driving environments, and simulate how the vehicles interact with each other.
However, these simulations rely heavily on the collected dataset, making it difficult to edit and expand the scenarios, resulting in the problem of scenario fragmentation.

Table \ref{tab1} provides a comparison of LimSim and other types of simulation systems in terms of functional coverage. 
Compared with three existing kinds of simulators, 
LimSim is able to simulate scenarios finely and efficiently, with a balance  between real-time and fidelity. It supports the generation of dynamic traffic scenario data and system testing for autonomous driving.

\section{System overview}
\subsection{System framework} % 系统总体架构、流程
As shown in Fig. \ref{fig:framework}, LimSim includes four modules to simulate long-term traffic scenarios, and the detailed descriptions of these modules are given below.

\paragraph{Multi-scenario road network construction} 
LimSim defines a standardized road network representation based on vector maps, which encompasses geometric and topological information. LimSim can cover multiple types of road scenarios such as intersections, roundabouts, and ramps. 
Users can construct the road network using tools such as SUMO or generate it directly from sensor data and expert knowledge. 
For long-term simulation, the route planning is carried out based on the standardized representation of the road network to construct road-level paths and determine feasible lanes for vehicles on each road section of the route. The scenario area of LimSim supports both fixed scene mode and hero mode (moving scenes following the ego vehicles). Vehicles in the scenario area have finer motion control to create a more realistic traffic environment.

\paragraph{Multi-source traffic flow generation} 
The multi-source traffic flow generation module in LimSim supports several baseline microscopic traffic models, such as car-following, lane-changing, and merging motions. 
It also supports customized scenarios using naturalistic driving datasets and relevant standards.
LimSim takes the continuity of traffic flow into account in both macroscopic and microscopic views, demonstrating compatibility and diversity while simulating real-world scenarios.

\paragraph{Multi-vehicle joint decision-making and planning} 
For controlled vehicles within the scenario area, LimSim adopts a hierarchical design that we call the \textit{PDP process}, including three steps: Prediction, Decision-making, and Planning. Such a design not only enables control for sophisticated vehicle behaviors but also matches the prevailing autonomous driving technology stack.
LimSim also adopts microscopic traffic models for predicting the vehicles outside the scenario area, acting as inputs for multi-vehicle joint decision-making and planning. 
% For controlled vehicles within the scenario area, joint behavior decision-making achieves various behaviors such as lane-keeping, lane-changing, acceleration and deceleration, and overtaking, reflecting the simulation efficiency and social features of vehicles. All controlled vehicles perform trajectory planning in parallel, taking into account costs such as safety, efficiency, and comfort. 

\paragraph{Multi-dimensional scenario analysis} LimSim captures information such as road network, vehicle properties, vehicle motion parameters, etc., throughout the simulation for scenario replay and data generation. It also provides online or offline analysis of the vehicle status during the simulation process, including indicators such as safety, comfort, energy consumption, and trajectory quality. 
The module supports a comprehensive evaluation of the simulation process including scenario complexity and risk.

The four modules of LimSim form an efficient and effective scenario-oriented simulation system that is used to test autonomous driving algorithms and facilitate transportation planning and policy-making. 
LimSim supports various driving behaviors and road scenarios while maintaining a realistic traffic environment and high data compatibility. % 这段话可删或调整

\subsection{User preference}

Users need to adapt the inputs of LimSim to complete traffic flow simulation tasks. 
According to the description of system modules, the input of LimSim must include the geometric dimensions of roads and the topological connections among them. Besides, traffic signal schemes or traffic sign information are optional inputs. 
LimSim supports vehicle control at signalized intersections and adjusts vehicle control strategies based on traffic sign information, such as adjusting the maximum vehicle speed based on road speed limits, or limiting whether vehicles are allowed to turn left or right based on traffic sign information. 
If only the road network information are provided, then LimSim can randomly generate demands and routes of vehicles for randomized simulation. We recommend the joint simulation of LimSim and SUMO, where necessary scenario information can be extracted from SUMO and relevant implementation scripts are provided. 
If users provide an instantaneous state of the scenario, including the moving vehicles in the road network, LimSim can take over and generate the subsequent motion paths of these vehicles. 
More generally, users can offer a collected scenario sequence, and LimSim can generate many simulation scenarios based on the original sequence. We have implemented this functionality based on the Waymo motion dataset. 

The output of LimSim includes real-time records of the simulation, which are available to users via the database, including road information, traffic control information, and vehicle trajectory information. Additionally, a visual user interface developed based on \text{Dear PyGui}\footnote[1]{\href{https://github.com/hoffstadt/DearPyGui}{https://github.com/hoffstadt/DearPyGui}} (a fast and powerful graphical user interface toolkit for Python) provides simulation visualization from the bird's eye view. 

% 可更改的配置信息
LimSim's interface allows users to replace algorithms used in PDP process with their own. Specifically, ego and other vehicles in the scenario are distinguished, and users can use self-developed algorithms to control the motion of the ego vehicle and then test the solutions' performance in the simulations. The scenario analysis module provides evaluation information for each simulation. Users can also extract the trajectories of all vehicles from the database to calculate custom performance metrics. 
For other vehicles in the scenario area, users can generate diverse scenarios by modifying the strategy parameters of the other vehicles. The customized parameters include the weights of the cost functions in trajectory planning, the social value orientation of vehicles, etc.

\begin{figure*}[htbp]
    \centering
    \includegraphics[width=\linewidth]{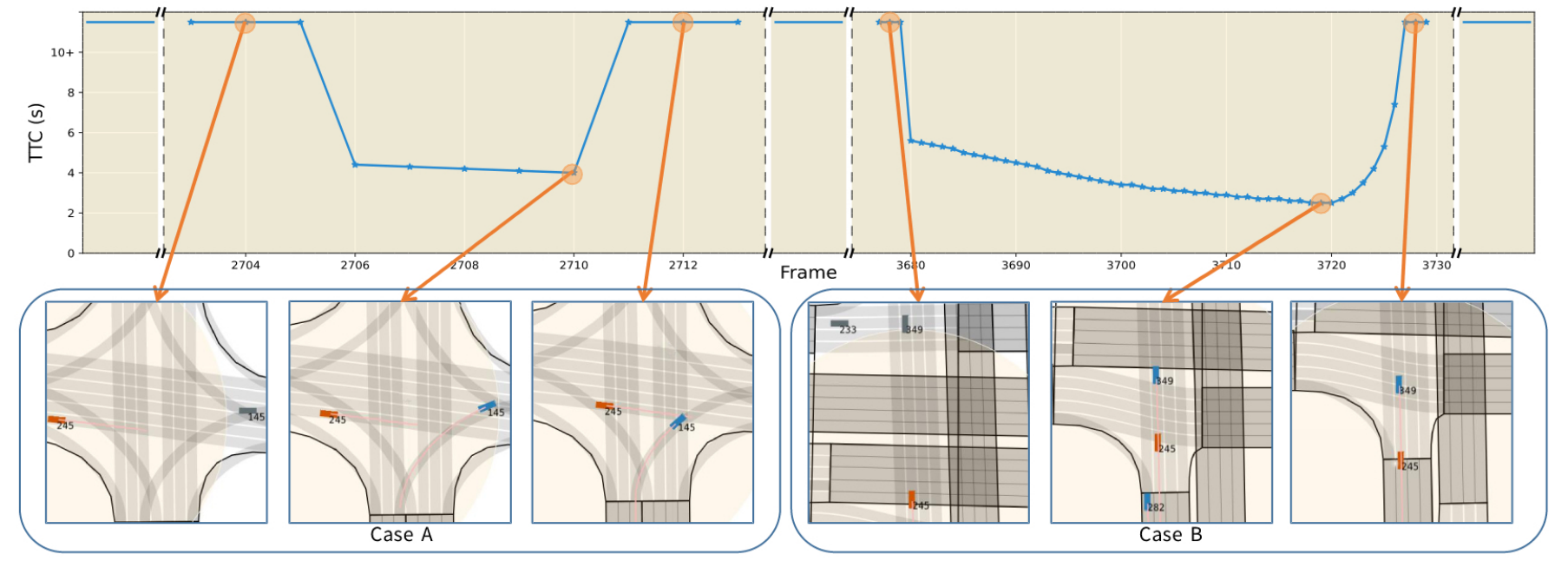}
    \caption{Extracting scenarios from long-term evaluation results. Case A describes a potential collision when a vehicle takes an unprotected left turn; Case B describes a potential rear-end collision where the following vehicle finally brakes to prevent crash. }
    \label{fig:scenario_extraction}
    \vspace{-15pt}
\end{figure*}

\section{Features}
\subsection{Long-term simulation and evaluation}

% 长时仿真描述，
% 案例分析，以一个具体的仿真来阐述
% -- 场景多样性（道路场景概述）、交通流多样性（行为统计）
% -- 评估方法，TTC公式等阐述，corner case判定

% 长时仿真，评估报告（），便于截取关键片段 @ maosong

% how to define key scenario? as not only corner cases are considered.

One important features of LimSim is extracting key scenarios through long-term simulation and evaluation.
Here we refer to key scenario as scenarios that provide a good specification for autonomous driving systems under design.
Classically, a large number of tests are required to estimate the performance of automated vehicles reliably due to rarity of key scenarios, which leads to high economic and time costs \cite{feng2023dense}.
To improve the efficiency of testing and validation, extracting key scenarios from driving data is critical.
Our strategy for extracting key scenarios  is ``One for all", that is, running one episode of simulation with all kinds of scenarios contained in it.
In LimSim, the city level map includes a range of complex scenarios such as roundabouts, intersections, motorways, etc. 
Different scenarios can be easily made available in testing phase. 
After setting up the environment,  customized algorithms can be sufficiently tested during one simulation episode.
When one episode of simulation is done, the key scenarios are extracted via evaluation results.

Most existing datasets lacks context of driving scenarios due to simplicity of scenario extraction methods. 
For example, Lyft dataset is partitioned by a fixed time frequency \cite{houston2021one}, which leads to discontinuity or incompleteness of a scenario. 
% Another strategy such as recording the braking signal, suffered from locating the start point of a scenario.
LimSim fills the gap by performing long-term scenario evaluation. 
We adopts the Time-to-collision (TTC), which is defined as ``The time required for two vehicles to collide if they continue at their present speed and on the same path''  \cite{Hayward1972NEARMISSDT}, as a measure of the severity of conflicts. 
Fig. \ref{fig:scenario_extraction} depicts extracting scenarios from the TTC curve.
In this simulation, the TTCs are above 10 seconds most of the time, indicating that those scenarios are ordinary. 
However, there are two segments with relatively short TTCs, both occurring in more critical scenes: a vehicle making an unprotected left turn and an emergency stop to avoid a rear-end collision.

Normally, only a few frames will be extracted to a key scenario, which may cause incompleteness. 
Benefit from the continuity of long-term simulation, LimSim improve this strategy by looking backward and forward. 
On the one hand, by adding frames before scenario begins, the states of tested and other vehicles are clearer and easier to simulate; on the other hand, adding frames after the ending of key scenario makes it possible to find a feasible trajectory.
Overall, the extra frames are added to the corner cases to serve the improvement and deployment of automated vehicles better.

\subsection{Multi-vehicle motion control}
% 决策规划联合框架更具体的描述（可以把mcts、分组等策略描述一下，画逻辑框图）
% 展示一些结果？

While popular microscopic traffic flow models partially simulate human driving habits, the social interaction behaviour between vehicles is still very homogeneous \cite{treiber2000Congested}. Nowhere is this more noticeable than when a vehicle is changing lanes and merging. 
These microscopic models do not take into account the vehicle's kinematic constraints and therefore fail to ensure smoothness and continuity of vehicle motion.
Thus, a more sophisticated control strategy is required for vehicles that have a direct impact on the ego vehicle, i.e. those in the scenario area.

\begin{figure}[tbp]
    \centering  
    \includegraphics[width=0.88\linewidth]{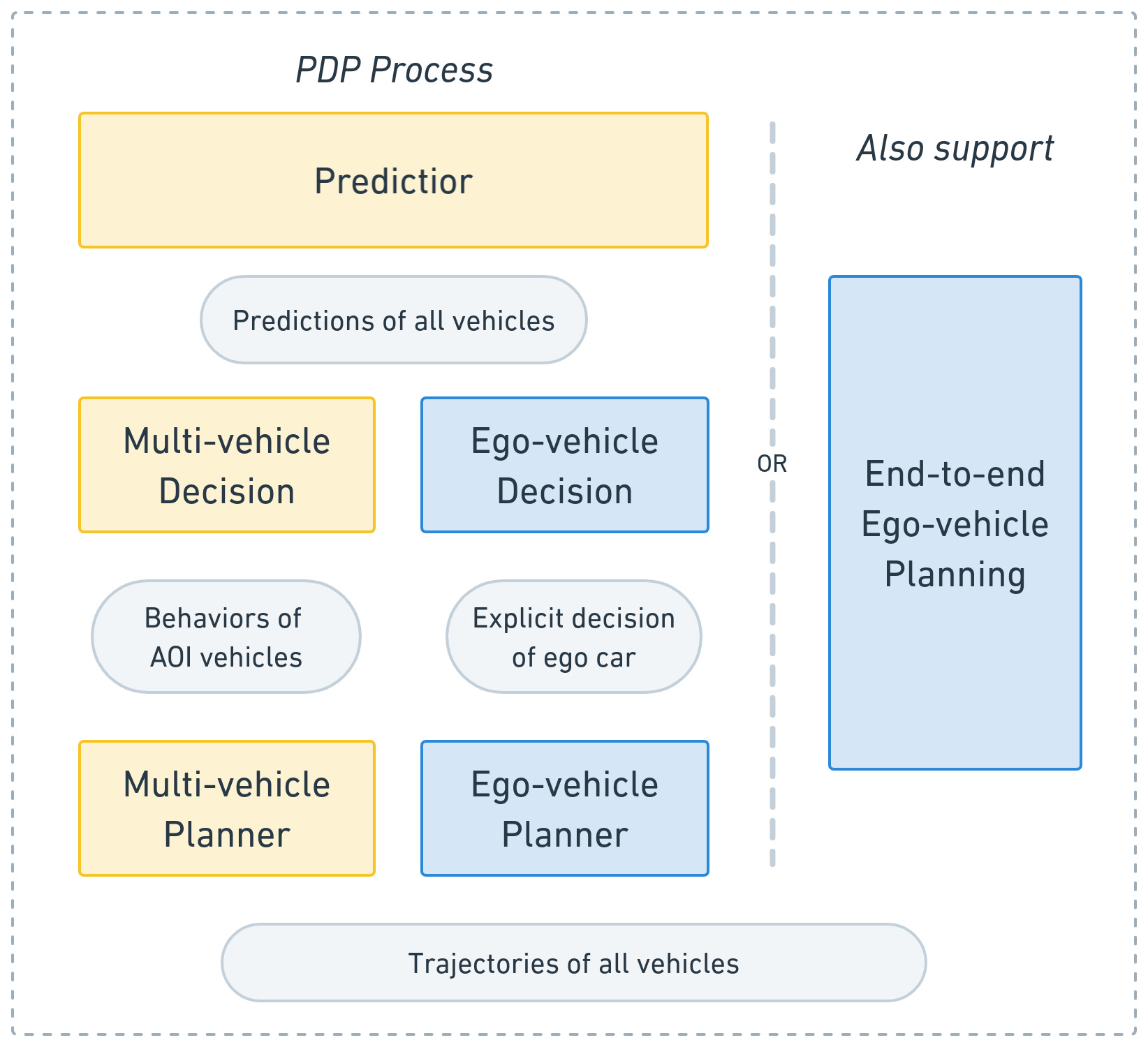}
    \caption{The PDP process in LimSim. The yellow boxes indicate the implemented procedures of LimSim, including multi-vehicle prediction, decision making and planning. The blue boxes indicate user-implemented procedures, focusing on ego-vehicle control.}
    \label{fig:pdp}
\vspace{-15pt}
\end{figure}

We adopts the hierarchical PDP process to perform motion control for vehicles within the scenario area. 
LimSim comes with an implementation of the PDP process modified from our previous work \cite{Wen2023} and outperforms traffic flow models in various metrics. 
LimSim also supports users' own implementation of part or all of the algorithm in the PDP process, allowing them to test their algorithm performance, as shown in Fig. \ref{fig:pdp}.

% Prediction 可以直接略过
% \subsubsection{Prediction}

% In the real world,  human drivers constantly need to estimate the intention and future movements of surrounding vehicles based on their experience.
% LimSim provides a prediction procedure to mimic such process. For those vehicles outside the scenario area, a uniformly accelerated motion model is adopted to predict its future trajectory, as only their historical information is available. For vehicles in the scenario area, the prediction algorithm uses more information, including routing details and previous decision results, to improve the accuracy of the predicted trajectories.
% This procedure predicts a trajectory for each vehicle in sensory range, and the user can choose how to use these predictions in subsequent decision and planning procedure.

\subsubsection{Decision-making}

Joint decision-making for multiple vehicle using Monte-Carlo Tree Search, or MCTS, has proven to allow for social vehicle interaction and generate traffic flows with diversity \cite{Li2022}. 
However, when simulating a busy intersection or a congested motorway, the large number of vehicles in the scenario area causes the search tree grows exponentially.  This results in a dramatic decrease of the computation efficiency and the success rate of decision-making algorithm.
Actually it's neither efficient nor realistic to require each vehicle  pays attention to every other vehicles' movements.
In LimSim, we employ a group-based MCTS decision-making algorithm for multiple vehicles  based on the idea of dividing the global vehicle interaction problem into a set of sub-problems with in each group.

Specifically, the decision-making procedure first calculate potential interactions between each two vehicles. For example, a cut-in vehicle and the vehicle behind it, and vehicles entering an intersection at the same time, both have some potential interactions. 
We divide all vehicles in the scenario area into several groups and make sure vehicles with potential interaction are assigned to the same group.
An intra-group MCTS decision-making algorithm is then performed for each group in turn and combined into a joint decision result for all vehicles in scenario area. 
The decision results of the previous groups are known and taken into account by the subsequent groups.

LimSim allows vehicles with different Social Value Orientations (SVOs) during the decision-making procedure \cite{Schwarting2019}. During MCTS, a vehicle considers not only the effect of the intended action on itself, i.e. reward to self, but also the risk of that action to other surrounding vehicles, i.e. penalty to others. 
Vehicles with different SVOs balance between these two factors in varying ways, resulting in radically divergent decision outcomes.

\subsubsection{Planning}

The planning algorithm with a parallel architecture fits well as a downstream procedure of the decision-making described above. 
The planner first attempts to generate a smooth and continuous path based on each vehicle's decision results. It also checks whether the path satisfies vehicle's kinematic constraints, including turning radius, speed/acceleration limits, etc.
If the above attempts fail and a vehicle is unable to follow its decision, the planner switches to a larger target state space to generate a valid trajectory, such as aborting a lane change or emergency braking.
Vehicles with different SVOs are also given different cost weights during the planning procedure, which further increases the diversity of vehicle trajectories.

\begin{figure}[tbp]
    \centering
     \includegraphics[width=0.8\linewidth]{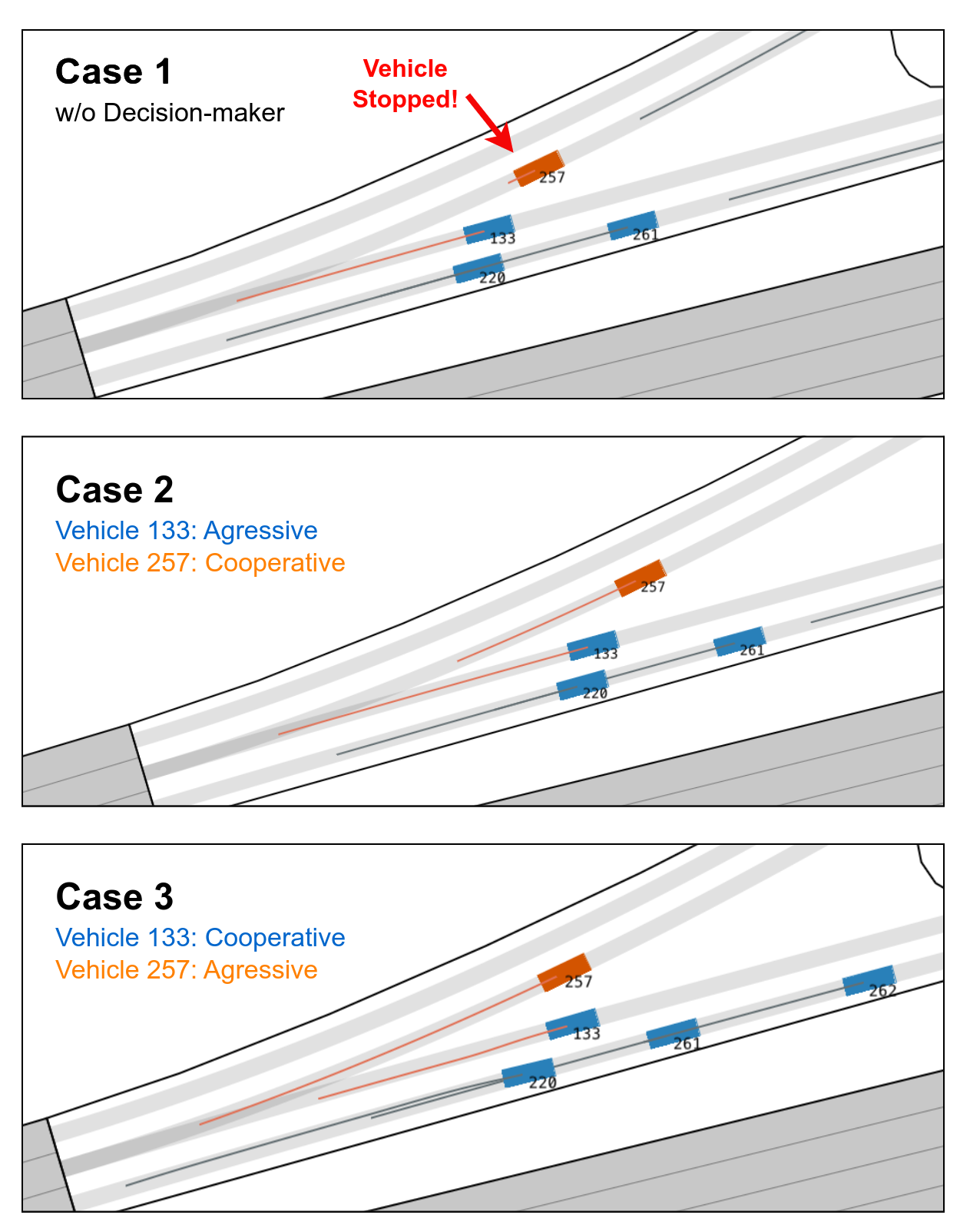}
      \caption{A merging scenario in highway. The line in the front of each vehicle indicates its trajectory for the next 3 seconds. In case 1 the decision-maker is disabled, causing ego vehicle performing an emergency stop. And in case 2 and case 3, vehicles possess different SVOs.}
      \label{fig:pdp_case}
\vspace{-15pt}
\end{figure}

\subsubsection{Case study}
% 分组的数量和每个组的车辆数会影响。
We demonstrate the performance of our PDP process in an on-ramp merging scenario from CitySim dataset \cite{citySim}. As shown in 
Fig. \ref{fig:pdp_case},  the vehicle with ID 257 on the ramp acts as the ego vehicle, and the vehicle with ID 133 on the mainline competes with it for the right of way. 

In case 1, the decision-making procedure is disabled. Due to a lack of foresighted long-term decisions, the ego vehicle missed the best time to merge onto mainline and thus had to brake urgently to avoid vehicle with ID 133.
Instead, in cases 2 and 3, the decision procedure is activated so that the dangerous situation never occurs.
In these two cases, vehicles possess different SVOs, where the aggressive style means that the vehicle is more selfish, and the cooperative style gives more consideration to other vehicles. As a result, the order of passage are changed.

\subsection{Interactive scenario reconstruction}

In a road environment, it would be very valuable for simulation if a large number of real scenarios could be collected and reconstructed.
Currently, there are two approaches to using real-world scenarios: scenario replay and scenario generation \cite{rempe2022generating}. 
In the approach of scenario replay, the simulation system provides decision-making and planning methods for the ego vehicle, while other vehicles in the scenario drive exactly as logged. Although easy to implement, the ego vehicle controlled by this method cannot interact with other vehicles and is limited to testing its responsiveness in very few scenarios.  
With regard to the scenario generation method, designed models are applied to plan trajectories for all vehicles in the log information from a certain moment until the end. Although it can achieve interaction between the ego vehicle and other vehicles, the subsequent vehicle movements may deviate completely from real scenarios due to insufficient model authenticity.

\begin{figure}[tbp]
    \centering
     \includegraphics[width=\linewidth]{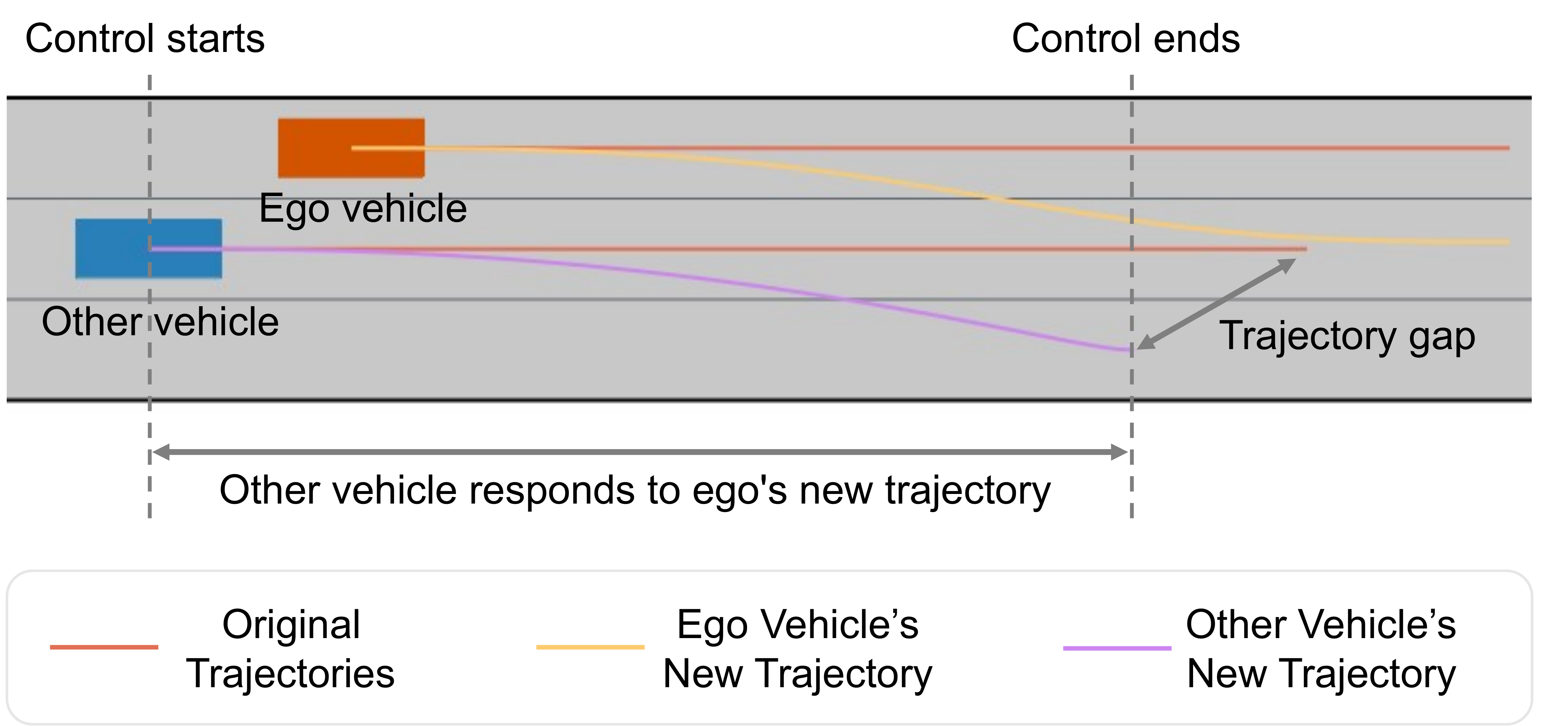}
      \caption{Illustration of interactive scenario reconstruction: When the ego vehicle adopts a new trajectory that affects another vehicle, the other vehicle must respond. At this point, the other vehicle is controlled by the LimSim to make an interactive action. When the other vehicle moves away from the ego vehicle, the control ends and the trajectory gap is considered for the other vehicle to return to its original trajectory.}
      \label{fig:interactiveReplay}
      \vspace{-15pt}
\end{figure}

LimSim offers the interactive scenario reconstruction as one of its core features, making the simulation scenarios as close to the real scenarios as possible.
In most cases, other vehicles in the simulation scenario will move according to their original trajectory in the log data. However, if other vehicles have to adjust their trajectories due to the impact of the ego vehicle, they need to be controlled. 
To achieve this goal, three problems have to be solved. 
Firstly, how to identify which vehicles need to be controlled by the simulation system? The original trajectories will be judged to see whether there is any conflict with other controlled vehicles by checking the multiple trajectories in the next few seconds. 
Secondly, how to determine the control period of the controlled vehicles? Based on the time length of the trajectory planning process, the controlled vehicles will obtain trajectories for next few seconds.  Once these vehicles reach the end points of the planned trajectories, they will be released from the control of the simulation system. 
Finally, after the vehicles are released from control, how to handle them? 
At the current time step, they will be destroyed if the vehicles do not have original trajectories in the log data.  If the vehicles still have their original trajectories, it is necessary to detect whether there is any conflict between the original trajectories and  trajectories of other vehicles in the scenario for the next few seconds. If there is no conflict, the vehicles will return to their original trajectories; otherwise,  vehicles will not be placed in the simulation scenario temporarily until all conflicts are resolved.

\subsection{Performance}
\begin{figure}[tbp]
    \centering
    \includegraphics[width=\linewidth]{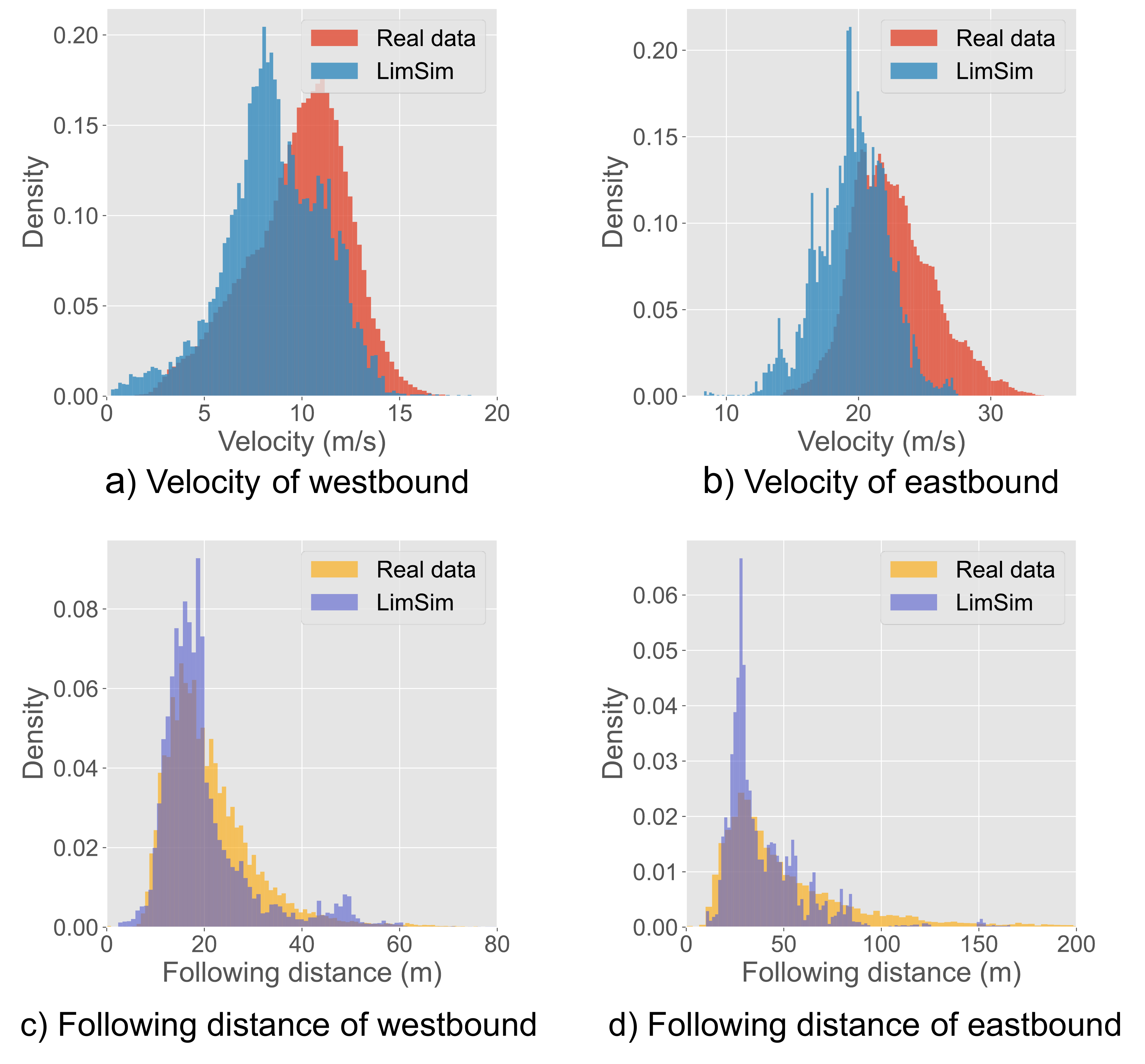}
    \caption{Comparison of velocity and following distance between LimSim and real data: In both the LimSim simulation and the real dataset, the speed distributions are close to the normal distribution, with similar mean and standard deviation values. The distributions of the following distance are also close to the Poisson distribution, with similar parameters between the simulated and real data. These results show that LimSim can effectively simulate human driving behavior and generate realistic traffic flow.}
    \label{fig:simulationCompare}
    \vspace{-15pt}
\end{figure}

% citysim 多长时间 多少辆车 我们通过对这些车轨迹的标定，提取出了每辆车在PDP过程中驾驶行为参数，
The generation of near-realistic traffic flows matters for a traffic simulator. LimSim can simulate human driving behavior and produce diverse traffic flows by adjusting the SVO parameters of the PDP process. 
To verify the performance of the simulated traffic flow generated by LimSim, we conduct a simulation based on the Freeway B dataset from CitySim \cite{citySim} and analyze the speed distribution and car-following distance distribution of the simulation results.  
The Freeway B dataset is a six-lane highway in both directions, containing two roads, a westbound road and an eastbound road.
The westbound road is relatively congested with a higher traffic volume than the the eastbound road. 
Simulation results are shown in Fig. \ref{fig:simulationCompare}.

The velocity distributions of the LimSim-generated traffic flows and the real data closely resemble the normal distribution with similar means and standard deviations. 
The mean speed of the LimSim-generated traffic flow on the westbound road is 8.57 m/s with a standard deviation of 2.57 m/s, while the mean speed of the real data is 9.81 m/s with a standard deviation of 2.58 m/s. 
The mean speed of the LimSim-generated traffic flow on the eastbound road is 19.57 m/s with a standard deviation of 2.67 m/s, while the mean speed of the real data is 22.62 m/s with a standard deviation of 3.11 m/s.  The average speed on the westbound road is lower than that on the eastbound road, which is consistent with the real dataset.
Two distributions of the car-following distance closely resembles the Poisson distribution, and the average distance in the eastbound direction is greater than that in the westbound direction, which is also consistent with the dataset. 
Overall, LimSim can simulate human driving behavior and provide traffic scenarios that are consistent with the dataset.

The experiment and demo videos of LimSim can be found on YouTube: \href{https://youtu.be/YR2A25v0hj4}{https://youtu.be/YR2A25v0hj4}.

% 跟车距离的分布则都近似于泊松分布，

\section{Conclusion}
In this paper, we introduce LimSim, the long-term interactive traffic simulator, which ensures the continuity of macroscopic and microscopic traffic flows and provides multi-type fine-grained dynamic traffic scenarios. 
LimSim adopts a closed-loop framework that supports a variety of driving behaviors in different scenarios while maintaining a realistic traffic environment and supporting multiple data formats.
LimSim performs a multi-dimensional real-time evaluation of the long-term simulation process to extract rare and challenging scenarios for further analysis. 
Besides, LimSim provides easy-to-use APIs to help users evaluate the performance of their own PDP solutions.

% 未来将持续扩展功能
We will open source our code and encourage community users to extend the current framework to support more personalised tasks, subject to license compliance. We will continue to maintain LimSim and continuously add useful features. We also plan to make LimSim compatible with various datasets to provide more realistic traffic scenarios for users involved in autonomous driving research.

\section*{Acknowledgments}
This work was supported by the Science and Technology Commission of Shanghai Municipality (Nos. 22YF1461400 and 22DZ1100102). 

\balance

\bibliographystyle{IEEEtran}
\bibliography{ref}

\begin{thebibliography}{10}
\providecommand{\url}[1]{#1}
\csname url@samestyle\endcsname
\providecommand{\newblock}{\relax}
\providecommand{\bibinfo}[2]{#2}
\providecommand{\BIBentrySTDinterwordspacing}{\spaceskip=0pt\relax}
\providecommand{\BIBentryALTinterwordstretchfactor}{4}
\providecommand{\BIBentryALTinterwordspacing}{\spaceskip=\fontdimen2\font plus
\BIBentryALTinterwordstretchfactor\fontdimen3\font minus
  \fontdimen4\font\relax}
\providecommand{\BIBforeignlanguage}[2]{{%
\expandafter\ifx\csname l@#1\endcsname\relax
\typeout{** WARNING: IEEEtran.bst: No hyphenation pattern has been}%
\typeout{** loaded for the language `#1'. Using the pattern for}%
\typeout{** the default language instead.}%
\else
\language=\csname l@#1\endcsname
\fi
#2}}
\providecommand{\BIBdecl}{\relax}
\BIBdecl

\bibitem{sagberg2015review}
F.~Sagberg, Selpi, G.~F. Bianchi~Piccinini, and J.~Engstr{\"o}m, ``A review of
  research on driving styles and road safety,'' \emph{Human factors}, vol.~57,
  no.~7, pp. 1248--1275, 2015.

\bibitem{zhang2021demystifying}
Y.~Zhang, W.~Jin, Z.~Xiong, Z.~Li, Y.~Liu, and X.~Peng, ``Demystifying
  interactions between driving behaviors and styles through self-clustering
  algorithms,'' in \emph{HCI in Mobility, Transport, and Automotive Systems},
  2021, pp. 335--350.

\bibitem{t1997preparing}
H.~Paul~'t, ``Preparing policy makers for crisis management: The role of
  simulations,'' \emph{Journal of Contingencies and Crisis Management}, vol.~5,
  no.~4, pp. 207--215, 1997.

\bibitem{teo2012evaluating}
J.~S. Teo, E.~Taniguchi, and A.~G. Qureshi, ``Evaluating city logistics measure
  in e-commerce with multiagent systems,'' \emph{Procedia-Social and Behavioral
  Sciences}, vol.~39, pp. 349--359, 2012.

\bibitem{daniel2021city}
M.~Daniel, R.~Dost{\'a}l, S.~Kozhevnikov, A.~Matyskov{\'a}, K.~Moudr{\'a},
  A.~M. Pereira, and O.~P{\v{r}}ibyl, ``City simulation software: Perspective
  of mobility modelling,'' in \emph{Smart City Symposium Prague (SCSP)}, 2021,
  pp. 1--7.

\bibitem{wang2022automatic}
S.-H. Wang, C.-H. Tu, and J.-C. Juang, ``Automatic traffic modelling for
  creating digital twins to facilitate autonomous vehicle development,''
  \emph{Connection Science}, vol.~34, no.~1, pp. 1018--1037, 2022.

\bibitem{argota2022getting}
J.~Argota S{\'a}nchez-Vaquerizo, ``Getting real: The challenge of building and
  validating a large-scale digital twin of {Barcelona}’s traffic with
  empirical data,'' \emph{ISPRS International Journal of Geo-Information},
  vol.~11, no.~1, p.~24, 2022.

\bibitem{schoner2018Simulation}
H.-P. Sch{\"o}ner, ``Simulation in development and testing of autonomous
  vehicles,'' in \emph{Internationales Stuttgarter Symposium: Automobil-und
  Motorentechnik}, 2018, pp. 1083--1095.

\bibitem{cameron1996PARAMICS}
G.~D. Cameron and G.~I. Duncan, ``{PARAMICS}—parallel microscopic simulation
  of road traffic,'' \emph{The Journal of Supercomputing}, vol.~10, pp. 25--53,
  1996.

\bibitem{fellendorf2010Microscopic}
M.~Fellendorf and P.~Vortisch, ``Microscopic traffic flow simulator {VISSIM},''
  \emph{Fundamentals of Traffic Simulation}, pp. 63--93, 2010.

\bibitem{casas2010Traffic}
J.~Casas, J.~L. Ferrer, D.~Garcia, J.~Perarnau, and A.~Torday, ``Traffic
  simulation with {Aimsun},'' \emph{Fundamentals of Traffic Simulation}, pp.
  173--232, 2010.

\bibitem{lopez2018Microscopic}
P.~A. Lopez, M.~Behrisch, L.~Bieker-Walz, J.~Erdmann, Y.-P. Fl{\"o}tter{\"o}d,
  R.~Hilbrich, L.~L{\"u}cken, J.~Rummel, P.~Wagner, and E.~Wie{\ss}ner,
  ``Microscopic traffic simulation using sumo,'' in \emph{International
  Conference on Intelligent Transportation Systems (ITSC)}, 2018, pp.
  2575--2582.

\bibitem{shah2017AirSim}
S.~Shah, D.~Dey, C.~Lovett, and A.~Kapoor, ``{AirSim}: High-fidelity visual and
  physical simulation for autonomous vehicles,'' in \emph{Field and Service
  Robotics: Results of the 11th International Conference}, 2018, pp. 621--635.

\bibitem{rong2020lgsvl}
G.~Rong, B.~H. Shin, H.~Tabatabaee, Q.~Lu, S.~Lemke, M.~Mo{\v{z}}eiko,
  E.~Boise, G.~Uhm, M.~Gerow, S.~Mehta \emph{et~al.}, ``{LGSVL} simulator: A
  high fidelity simulator for autonomous driving,'' in \emph{IEEE International
  Conference on Intelligent Transportation Systems (ITSC)}, 2020, pp. 1--6.

\bibitem{dosovitskiyCARLA}
A.~Dosovitskiy, G.~Ros, F.~Codevilla, A.~Lopez, and V.~Koltun, ``{CARLA}: An
  open urban driving simulator,'' in \emph{Conference on Robot Learning
  (CoRL)}, 2017, pp. 1--16.

\bibitem{li2022MetaDrive}
Q.~Li, Z.~Peng, L.~Feng, Q.~Zhang, Z.~Xue, and B.~Zhou, ``{MetaDrive}:
  Composing diverse driving scenarios for generalizable reinforcement
  learning,'' \emph{IEEE Transactions on Pattern Analysis and Machine
  Intelligence}, vol.~45, no.~3, pp. 3461--3475, 2022.

\bibitem{bergamini2021SimNet}
L.~Bergamini, Y.~Ye, O.~Scheel, L.~Chen, C.~Hu, L.~Del~Pero, B.~Osi{\'n}ski,
  H.~Grimmett, and P.~Ondruska, ``{SimNet}: Learning reactive self-driving
  simulations from real-world observations,'' in \emph{IEEE International
  Conference on Robotics and Automation (ICRA)}, 2021, pp. 5119--5125.

\bibitem{sun2022InterSim}
Q.~Sun, X.~Huang, B.~C. Williams, and H.~Zhao, ``{InterSim}: Interactive
  traffic simulation via explicit relation modeling,'' in \emph{IEEE/RSJ
  International Conference on Intelligent Robots and Systems (IROS)}, 2022, pp.
  11\,416--11\,423.

\bibitem{feng2022trafficgen}
L.~Feng, Q.~Li, Z.~Peng, S.~Tan, and B.~Zhou, ``{TrafficGen}: Learning to
  generate diverse and realistic traffic scenarios,'' \emph{ArXiv preprint
  arXiv:2210.06609}, 2022.

\bibitem{owen2000Traffic}
L.~E. Owen, Y.~Zhang, L.~Rao, and G.~McHale, ``Traffic flow simulation using
  {CORSIM},'' in \emph{Winter Simulation Conference Proceedings}, vol.~2, 2000,
  pp. 1143--1147.

\bibitem{xu2019Automated}
J.~Xu, Q.~Luo, K.~Xu, X.~Xiao, S.~Yu, J.~Hu, J.~Miao, and J.~Wang, ``An
  automated learning-based procedure for large-scale vehicle dynamics modeling
  on {Baidu} {Apollo} platform,'' in \emph{IEEE/RSJ International Conference on
  Intelligent Robots and Systems (IROS)}, 2019, pp. 5049--5056.

\bibitem{craighead2007Survey}
J.~Craighead, R.~Murphy, J.~Burke, and B.~Goldiez, ``A survey of commercial \&
  open source unmanned vehicle simulators,'' in \emph{IEEE International
  Conference on Robotics and Automation (ICRA)}, 2007, pp. 852--857.

\bibitem{carpin2007usarsim}
S.~Carpin, M.~Lewis, J.~Wang, S.~Balakirsky, and C.~Scrapper, ``{USARSim}: a
  robot simulator for research and education,'' in \emph{IEEE International
  Conference on Robotics and Automation (ICRA)}, 2007, pp. 1400--1405.

\bibitem{michel2004cyberbotics}
O.~Michel, ``Cyberbotics ltd. webots™: professional mobile robot
  simulation,'' \emph{International Journal of Advanced Robotic Systems},
  vol.~1, no.~1, p.~5, 2004.

\bibitem{laue2006simrobot}
T.~Laue, K.~Spiess, and T.~R{\"o}fer, ``{SimRobot}--a general physical robot
  simulator and its application in robocup,'' in \emph{RoboCup 2005: Robot
  Soccer World Cup IX}, 2006, pp. 173--183.

\bibitem{juliani2018unity}
A.~Juliani, V.-P. Berges, E.~Teng, A.~Cohen, J.~Harper, C.~Elion, C.~Goy,
  Y.~Gao, H.~Henry, M.~Mattar \emph{et~al.}, ``Unity: A general platform for
  intelligent agents,'' \emph{ArXiv preprint arXiv:1809.02627}, 2018.

\bibitem{pilz2019Development}
C.~Pilz, G.~Steinbauer, M.~Schratter, and D.~Watzenig, ``Development of a
  scenario simulation platform to support autonomous driving verification,'' in
  \emph{IEEE International Conference on Connected Vehicles and Expo (ICCVE)},
  2019, pp. 1--7.

\bibitem{koenig2004design}
N.~Koenig and A.~Howard, ``Design and use paradigms for gazebo, an open-source
  multi-robot simulator,'' in \emph{IEEE/RSJ International Conference on
  Intelligent Robots and Systems (IROS)}, vol.~3, 2004, pp. 2149--2154.

\bibitem{feng2023dense}
S.~Feng, H.~Sun, X.~Yan, H.~Zhu, Z.~Zou, S.~Shen, and H.~X. Liu, ``Dense
  reinforcement learning for safety validation of autonomous vehicles,''
  \emph{Nature}, vol. 615, no. 7953, pp. 620--627, 2023.

\bibitem{houston2021one}
J.~Houston, G.~Zuidhof, L.~Bergamini, Y.~Ye, L.~Chen, A.~Jain, S.~Omari,
  V.~Iglovikov, and P.~Ondruska, ``One thousand and one hours: Self-driving
  motion prediction dataset,'' in \emph{Conference on Robot Learning}.\hskip
  1em plus 0.5em minus 0.4em\relax PMLR, 2021, pp. 409--418.

\bibitem{Hayward1972NEARMISSDT}
J.~C. Hayward, ``Near-miss determination through use of a scale of danger,''
  \emph{Highway Research Record}, 1972.

\bibitem{treiber2000Congested}
M.~Treiber, A.~Hennecke, and D.~Helbing, ``Congested traffic states in
  empirical observations and microscopic simulations,'' \emph{Physical review
  E}, vol.~62, no.~2, p. 1805, 2000.

\bibitem{Wen2023}
L.~Wen, P.~Cai, D.~Fu, S.~Mao, and Y.~Li, ``Bringing diversity to autonomous
  vehicles: An interpretable multi-vehicle decision-making and planning
  framework,'' \emph{ArXiv preprint arXiv:2302.06803}, 2023.

\bibitem{Li2022}
C.~Li, T.~Trinh, L.~Wang, C.~Liu, M.~Tomizuka, and W.~Zhan, ``Efficient
  game-theoretic planning with prediction heuristic for socially-compliant
  autonomous driving,'' \emph{IEEE Robotics and Automation Letters}, vol.~7,
  no.~4, pp. 10\,248--10\,255, 2022.

\bibitem{Schwarting2019}
W.~Schwarting, A.~Pierson, J.~Alonso-Mora, S.~Karaman, and D.~Rus, ``Social
  behavior for autonomous vehicles,'' \emph{Proceedings of the National Academy
  of Sciences}, vol. 116, no.~50, pp. 24\,972--24\,978, 2019.

\bibitem{citySim}
O.~Zheng, M.~Abdel-Aty, L.~Yue, A.~Abdelraouf, Z.~Wang, and N.~Mahmoud,
  ``{CitySim}: A drone-based vehicle trajectory dataset for safety oriented
  research and digital twins,'' \emph{ArXiv preprint arXiv:2208.11036}, 2022.

\bibitem{rempe2022generating}
D.~Rempe, J.~Philion, L.~J. Guibas, S.~Fidler, and O.~Litany, ``Generating
  useful accident-prone driving scenarios via a learned traffic prior,'' in
  \emph{Proceedings of the IEEE/CVF Conference on Computer Vision and Pattern
  Recognition (CVPR)}, 2022, pp. 17\,305--17\,315.

\end{thebibliography}

\end{document}